\documentclass{article}
\usepackage[english]{babel}
\usepackage[margin=1in]{geometry}

\usepackage{amsmath}
\usepackage{graphicx}
\usepackage[colorlinks=true, allcolors=blue]{hyperref}
\usepackage{authblk}

\title{Effect on spectral purity due to on-chip temporal manipulation of the pump}
\author[1]{Lukas Rodda}
\author[1,2]{Ben M. Burridge}
\author[1]{Jorge Barreto}
\author[1,*]{Imad I. Faruque}

\affil[1]{Quantum Engineering Technology Labs, University of Bristol, Nanoscience and Quantum Information Building, Tyndall Avenue, Bristol BS8 1FD, United Kingdom}
\affil[2]{Quantum Engineering Centre for Doctoral Training, Centre for Nanoscience \& Quantum Information, University of Bristol, Bristol, United Kingdom}
\affil[*]{imad.faruque@bristol.ac.uk}
\begin{document}
\maketitle

\begin{abstract}
Photonic Integrated Circuits (PIC)s are a promising contender for quantum information technologies. The spectral purity of photons is one of the key attributes of PIC photon-pair sources. The dual-pulse pump manipulation technique \cite{Christensen:18} showed $>$99\% purity in ring-resonator photon-pair sources. 
Here, we have developed a PIC to shape a pulse into dual, triple and quadruple pulses and investigated the effect of these pump pulse configurations on the purity. Our results show that more complex configurations over dual-pulse do not result in comparatively higher purity but allow accurate control over choosing arbitrary values of the purity.
\end{abstract}

\section{Introduction}
Quantum information is the foundation for many strategic technologies such as quantum computing and communication \cite{ion,YUAN20101,Wang2020,Huang2020}. Integrated photonics has shown to be a promising platform to scale-up and more stable quantum experiments while benefiting from well developed fabrication techniques \cite{Chen2011}.
Using photons as quantum bits (qubits) requires a source of single photons, such as a heralded single-photon source. Recently published results present various methods to improve the spectral purity of the photon sources, which is crucial for high-fidelity gate operation \cite{Caspani2017}. For example, complex resonator designs \cite{burridge2023integrate}, or temporal manipulations of the pulsed pumps \cite{Christensen:18, Burridge:20} are two of the ways to achieve near unity purity, an essential characteristic of photon pair sources for heralding.. 

\begin{figure}[t]
    \centering
    \includegraphics[width=0.5\linewidth]{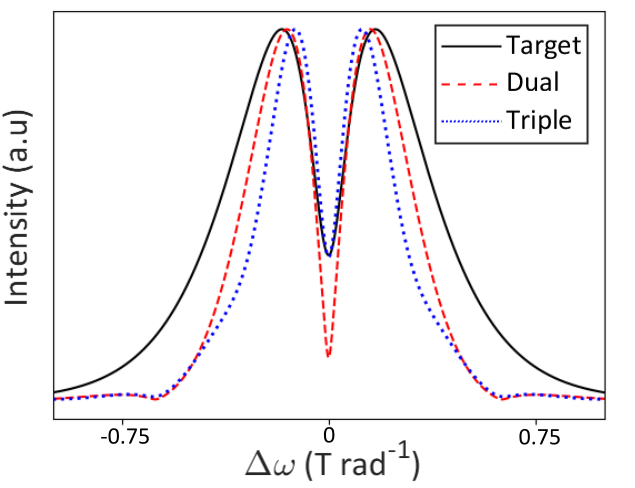}
    \caption{Target pulse for completely pure JSA, compared to dual and triple pulse pumps. The dual pulse shown has a splitting ratio 0.55 and a delay of 10ps, the best performing case. The triple pulse has splitting ratios of 0.8 and 0.8, and a 10ps delay between each pulse}. 
    \label{fig:sim}
\end{figure}

Nonlinear light-matter interaction mechanisms such as spontaneous four-wave mixing (SFWM) are often used in photonic devices in integrated photonics platforms (e.g., silicon-on-insulator platform) to generate photon-pairs, where two pump photons are converted into a signal and idler photon. A ring-resonator is one such structure that enhances the production of photon-pairs with high brightness and high purity \cite{Caspani2017}. The spectral purity of this source is often quantified using joint spectral amplitude (JSA) of emitted signal and idler photons. This is given by \cite{Helt:10}
\begin{equation}\label{eq:jsa}
    \Phi(\omega_s,\omega_i) = \int d\omega_{p}d\omega_{p'}\hspace{4pt}\alpha(\omega_{p})\alpha(\omega_{p'})\phi(\omega_{p},\omega_{p'},\omega_s,\omega_i) L(\omega_{p})L(\omega_{p'})L(\omega_s)L(\omega_i)
\end{equation}
where $\alpha(\omega)$ is the pump envelope, $\phi$ is the phase matching term, $L(\omega)$ is the Lorentzian field enhancement terms arising from the ring resonances, with pump, signal and idler frequencies: $\omega_p$, $\omega_s$ and $\omega_i$. The pump is spectrally confined by the ring resonance at the pump frequencies, which causes correlations in the photon pairs. Theoretically the purity of a photon pair state generated in a ring is limited to 0.92 for a single pulse pump \cite{Helt:10}. However, by manipulating the phase and temporal spectrum of the pump, the effect of the resonance at the pump frequencies can be removed, leading to a separable pair state and arbitrarily high purity \cite{Christensen:18}.
\begin{equation}\label{eq:target}
    \alpha_{target} \propto L(\omega_p)^{-1}\times \alpha_{single}
\end{equation}
This ideal target pump can be approximated using a dual pulse (Fig.~\ref{fig:sim}a), of two $\pi$ phase-shifted pulses, which has been theoretically shown to improve purity past 0.99 \cite{Christensen:18}, and was demonstrated experimentally using fibre-optics \cite{Burridge:20}, however pump configurations were only stable for approximately 1h and tuning ranges of the parameters were limited.
\begin{figure*}[t]
\centering\includegraphics[width=\linewidth]{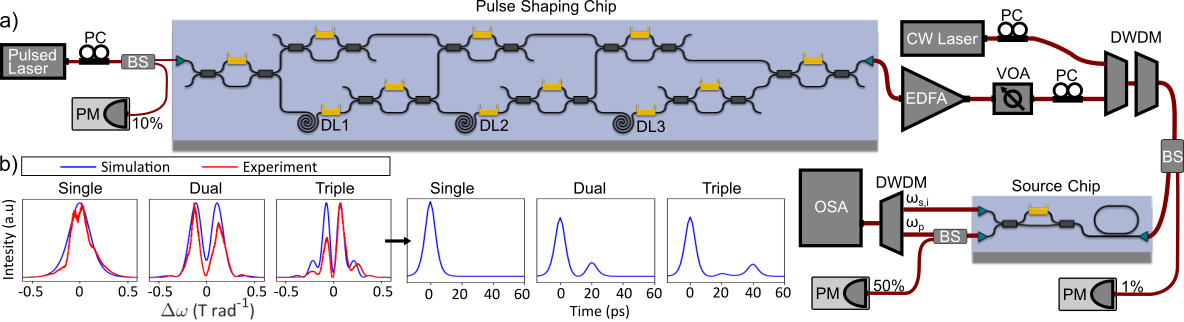}
\caption{a) Experimental set-up used to collect purity measurements for different pulse pump shapes. Pulses are created in the first chip using three delay lines (DL), then combined with a stimulating signal field from a continuous wave (CW) laser, to generate photon pairs in a ring resonator source on the second chip. An optical spectrum analyser (OSA) then measures the intensity of generated idler photons. Also used are: polarisation controllers (PC), beam splitters (BS), powermeters (PM), an erbium-doped fibre amplifier (EDFA), a variable optical attenuator (VOA) and dense-wavelength division multiplexers (DWDM). b) Single, dual and triple pulses, from experimental and simulated data, where experimental data was collected by the OSA after setting the parameters on-chip, and simulated data from models using the same parameters. For the dual pulse $\eta=0.8$ and $\Delta\tau=20ps$ (modelled with equation \ref{eq:dual}). For the triple pulse $\eta_1, \eta_2 = 0.8,0.2$ and $\Delta\tau_1 = 20ps$, $\Delta\tau_2 = 40ps$ (modelled with equation \ref{eq:tripl}). An inverse Fourier transform was used to plot the simulated pulses in time domain.}
\label{fig:setup}
\end{figure*}
Here we present simulated and experimental results which explore alternate pump profiles, such as a pump containing three pulses, to therefore match, improve or provide more control over the purity of a dual pulse pump. We used a PIC for pulse pump manipulation. This allowed multiple pump pulse configurations compared to the literature \cite{Burridge:20, Christensen:18}, and also allowed for accurate control over parameters while guaranteeing the phase stability of the composite pump pulse as parameters are changed. The pump can then be used with any conventional resonator source, allowing for easy integration into new or existing photonic experiments.

\section{Methods}

The set up shown in Fig.~\ref{fig:setup}a was used to create various pulse pump configurations and record the spectral purity of our photon pair source. Our experiment uses two 220~nm thickness silicon-on-insulator chips. We used a pulsed laser (PriTEL FFL) which produces $sech^2$ shaped single pulses at 1550nm wavelength with 41~GHz Full-Width Half Maximum (FWHM) and 50~MHz repetition rate. Such a pulse pump $\alpha_{single}$ is described using the following equation
\begin{equation}
    \alpha_{single} = sech^2((\omega-\omega_0)/\sigma)
    \label{eq:single}
\end{equation}
where, $\omega$ denotes angular frequency, $\omega_0$ denotes central frequency and $\sigma$ denotes FWHM of the pulse. These pulses were coupled to the first chip, denoted as the pulse shaping chip in the figure, through a polarisation controller to optimise the coupling into the chip using a vertical grating coupler, as well as a beam splitter as a 10\% tap to monitor the input power in the chip. 
Each single pulse is split into two by a Mach-Zehnder interferometer (MZI). The MZI is controlled by a thermo-optic phase shifter to have the desired splitting ratio $\eta_1$. The two pulses are then temporally separated using a fixed delay ($\Delta\tau_1$) line of 20~ps ($\Delta\tau_1=$~20~ps) and then recombined using another MZI. The relative phase difference ($\phi_1$) between the two pulses can also be set using another thermo-optic phase shifter. At this point in the circuit, the resulting pulse configuration can be expressed by Eq.~\ref{eq:dual} which can be denoted as a dual pulse. 
\begin{equation}
    \alpha_{dual} = [\sqrt{\eta_1} + \sqrt{1-\eta_1}\exp{(-i\Delta\tau_1(\omega-\omega_0)+i\phi_1)}]\times\alpha_{single}
    \label{eq:dual}
\end{equation}
The same process continues in series twice more before the resulting pulse is output from the chip, allowing a pulse configuration of up to four pulses to be created with arbitrary splitting ratios and phases, and with fixed 20~ps time delays between them. Thus, a triple pulse, with another splitting ratio ($\eta_2$), time delay ($\Delta\tau_2$) and phase ($\phi_2$), can be expressed as
\begin{multline}
    \alpha_{triple} = [\sqrt{\eta_1}+\sqrt{(1-\eta_1)\eta_2}\exp((-i\Delta\tau_1(\omega-\omega_0)+i\phi_1)\\
    +\sqrt{(1-\eta_1)(1-\eta_2)}\exp(-i\Delta\tau_2(\omega-\omega_0)+i\phi_2)]\times\alpha_{single}
    \label{eq:tripl}
\end{multline}
Where all delays and phases are relative to the first pulse maximum.

A few configurations are plotted in Fig.~\ref{fig:setup}b, where the single pulse Eq.~\ref{eq:single} was numerically fit to the corresponding output of the chip, and the dual and triple pulse Eqs.~\ref{eq:dual} and \ref{eq:tripl} were fit using the same values for $\eta$, $\Delta\tau$ and $\phi$ as were set on chip. Once the desired pulse configuration was acquired, it was amplified using an EDFA to be used as a strong laser pump for stimulated emission tomography (SET) in the second chip \cite{set}. 


The second chip contains a racetrack resonator source which can produce photon-pairs through a nonlinear light-matter interaction through SFWM. To perform SET, we have combined our pump laser with a continuous wave (CW) laser (Yenista Tunics) as a seed field using a DWDM (Fig.~\ref{fig:setup}(a)). The pump and the seed field produce a stimulated idler field in the resonator. A variable optical attenuator was used to bring the power to an arbitrary level, and an extra DWDM was used to filter the noise after the EDFA. This was then coupled into the second chip, with a 1\% tap to monitor the combined power and two PCs to optimise the coupling. 
 
The racetrack resonator is followed by an asymmetric-MZI that removes the pump field from the signals. After the chip, the pump was monitored using another tap. The pump and the stimulated idler fields were combined with a final DWDM which filtered out the seed field. The intensity of the generated idler photons was measured using a (Finisar WaveAnalyser) optical spectrum analyser (OSA).
A single pulse JSI was collected using SET first by configuring the pulse shaper circuit so that the original pulse wasn't split. Then a triple pulse was created by splitting a single pulse twice with a ratio of $\eta_1 = \eta_2 = 0.8$, and the JSI was collected using SET.

The JSA is approximated using experimentally measured $\sqrt{JSI}$. Alternatively we simulate the JSA numerically in MATLAB with Eq.~\ref{eq:jsa} and the corresponding pulse shape ($\alpha$). We then calculate the purity for each pulse pump configuration through a Schmidt decomposition of the JSA \cite{Helt:10}. We have also simulated the in-resonator field spectral distribution given by $A_p(\omega) = L(\omega - \omega_p)\times \alpha_{pump}((\omega -\omega_p))$, and the corresponding temporal distribution using Fourier transformations.
\section{Results and discussion}
The pulse shaper circuit was characterised by creating single, dual and triple pulse pumps with varying splitting ratios, phases and delays. These matched up well with our model (Fig.~\ref{fig:setup}b), showing the chip is capable of a fine level of control. The single, dual and triple pulse pumps were modelled with Eqs.~\ref{eq:single}, \ref{eq:dual} and \ref{eq:tripl}.

\begin{figure}[ht!]
\centering\includegraphics[width=0.6\linewidth]{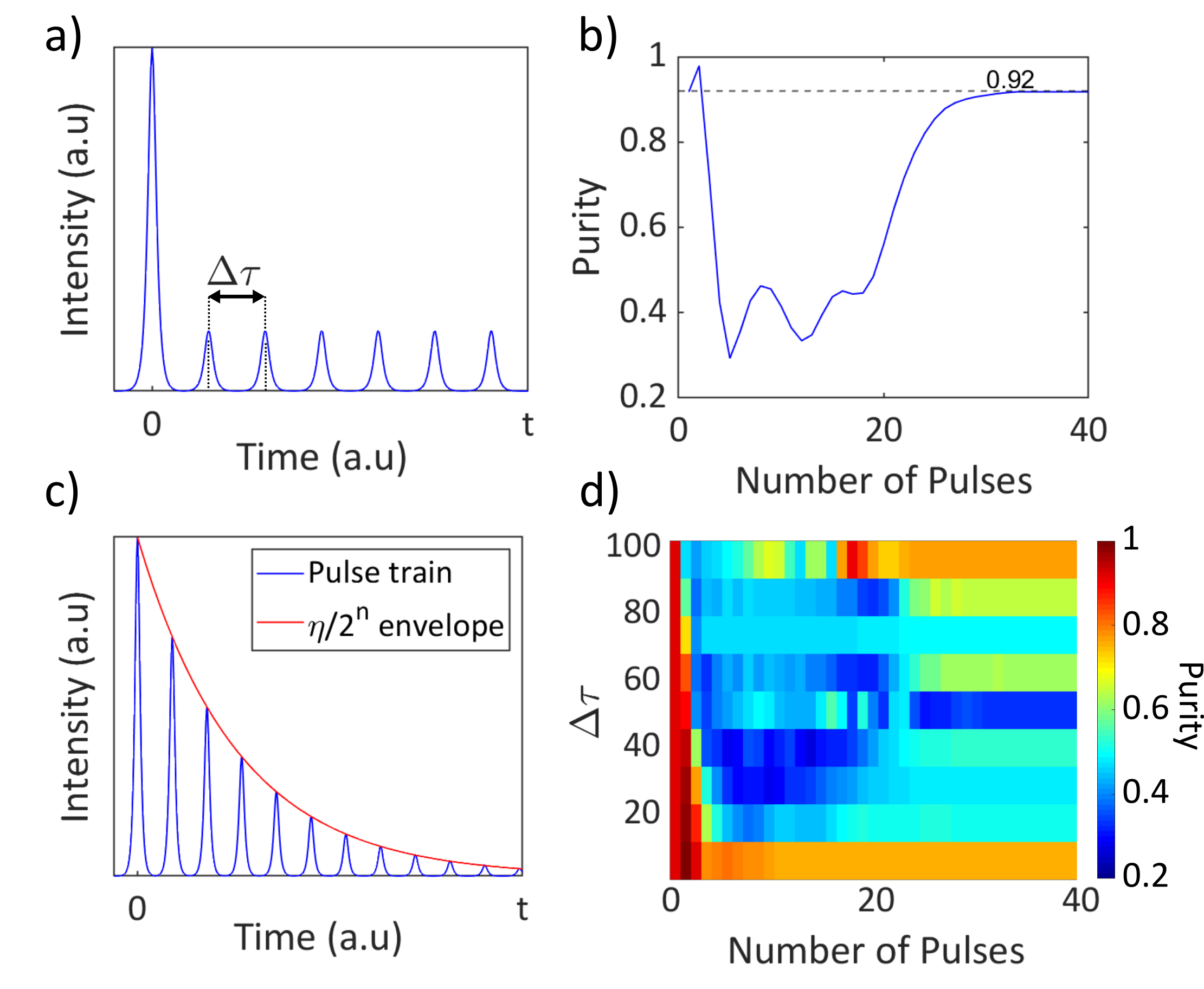}
\caption{a) An example of a pulse train used in the $n$ pulse pump simulations. Here the amplitude is constant after an initial stronger pulse, with a time delay $\Delta\tau$ between each pulse b) Purity from simulated JSAs as the number of pulses increases, for the constant pulse train. c) Pulse train where amplitude decreases according to a splitting ratio $\eta$, for each pulse $n$, as shown by the envelope. d) Purity as the number of pulses increases, with different delays ($\Delta\tau$) between them. Here the pulse train amplitudes decreased according to a splitting ratio $\eta = 0.55$, as shown in c).}
\label{fig:heat}
\end{figure}

An $n$ pulse pump was modelled to investigate the purity of a pump with an arbitrary number of pulses. A train of pulses was constructed by adding $n$ single pulses together, with a constant time delay between them and all $\pi$ out of phase with respect to the first pulse. Then two different configurations, in terms of the pulse amplitudes, were used. First, the pulse train was set so that after an initial pulse, each next pulse had a constant, lower amplitude as shown in Fig.~\ref{fig:heat}a. Then a pulse train was set where the amplitude of each subsequent $n$ pulse was reduced using a splitting ratio $\eta/2^n$ as shown in Fig.~\ref{fig:heat}c, similar to how the pulse shaper circuit operates. 
Both of these configurations showed similar results (Figs.~\ref{fig:heat}b, d). No improvement in purity over the dual pulse case was seen as the number of pulses increased. Furthermore, only a few cases exceed the 0.92 purity limit, with the first configuration only plateauing at the limit as the pulse train approximated a single pulse. Since these configurations have fixed time delays and phases, to find higher purity solutions, we focus on the triple pulse configuration (Eq.~\ref{eq:tripl}) where the splitting ratios and phases can be varied arbitrarily to explore a larger parameter space to illustrate any benefits over dual pulse case.

\begin{figure}
    \centering
    \includegraphics[width=0.6\linewidth]{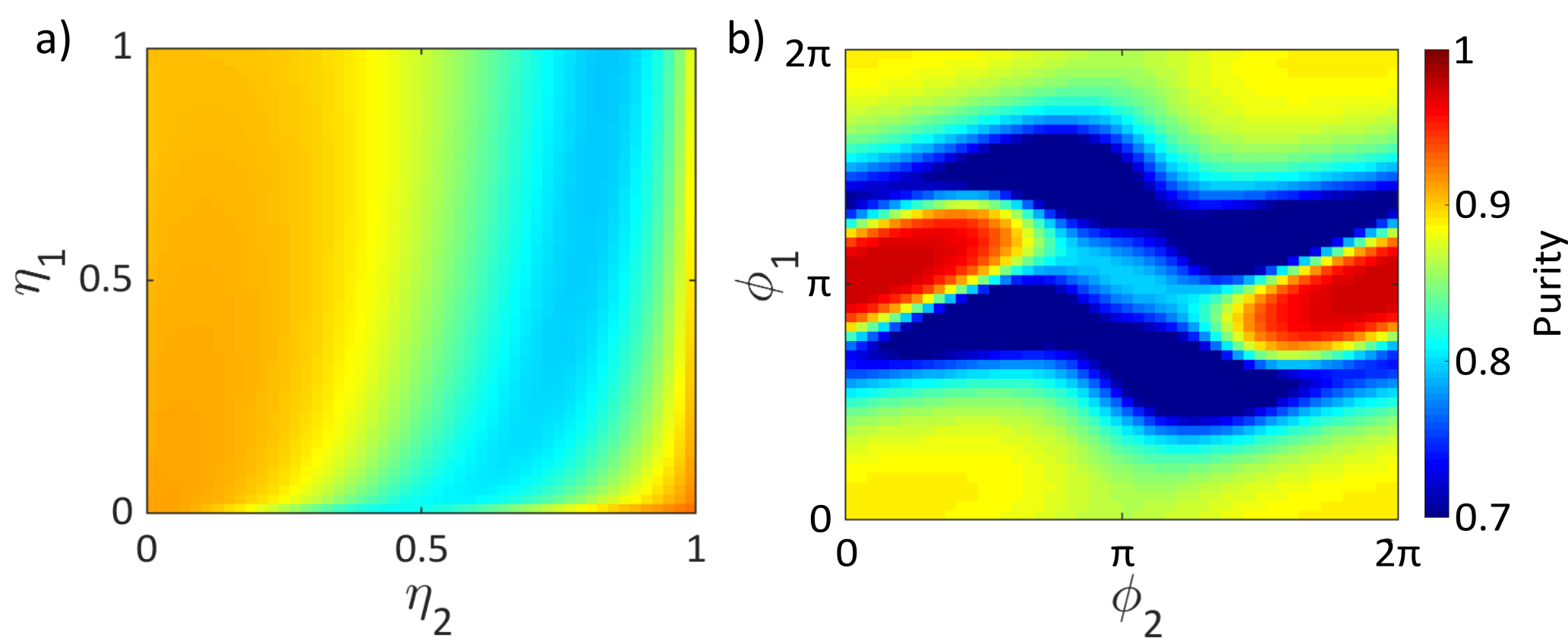} 
    \caption{a) Purity as splitting ratios $\eta_1$ and $\eta_2$ are varied, for a triple pulse with $\Delta\tau_1$ and $\Delta\tau_2$ set to 20ps and 40ps respectively, and $\phi_1$, $\phi_2$ both set to $\pi$. b) Purity as the phase of the second and third pulse is varied relative to the first. Here $\eta_1$ and $\eta_2$ are both 0.8, with the same delays as previously.}
    \label{fig:triple sim}
\end{figure}

First the two splitting ratios ($\eta_{1,2}$) were varied while keeping the delays ($\Delta\tau_{1,2}$) and phases ($\phi_{1,2}$) constant at 20ps and $\pi$ respectively. Then $\eta_1$ and $\eta_2$ were both fixed to 0.8, and the phases were varied, keeping the delays the same. The parameters were scanned, simulating the JSA for each configuration, creating two heatmaps shown in Fig.~\ref{fig:triple sim}.
 The results show (Fig.~\ref{fig:heat}) that the purity does not reach higher than the best dual pulse case, and only a few configurations perform better than the single pulse limit. For simulations of the dual pulse (Eq.~\ref{eq:dual}), the best performing configuration reaches 0.999 purity, with a large set of the parameter space going above 0.99 \cite{Christensen:18}. On the other hand, almost all of the space for the triple pulse gives purity at or below the 0.92 limit, with only a small area in the phase simulation reaching higher than 0.95. However, this shows that a wide range of purity values can be selected simply by changing the phase and splitting ratios, with high tolerance. This suggests more control over the dual pulse, where changing the time delay is required for similar control, which is more complicated to achieve experimentally and can only be achieved in limited discrete time steps on a photonic chip.
\begin{figure}[h!] 
\centering\includegraphics[width=0.6\linewidth]{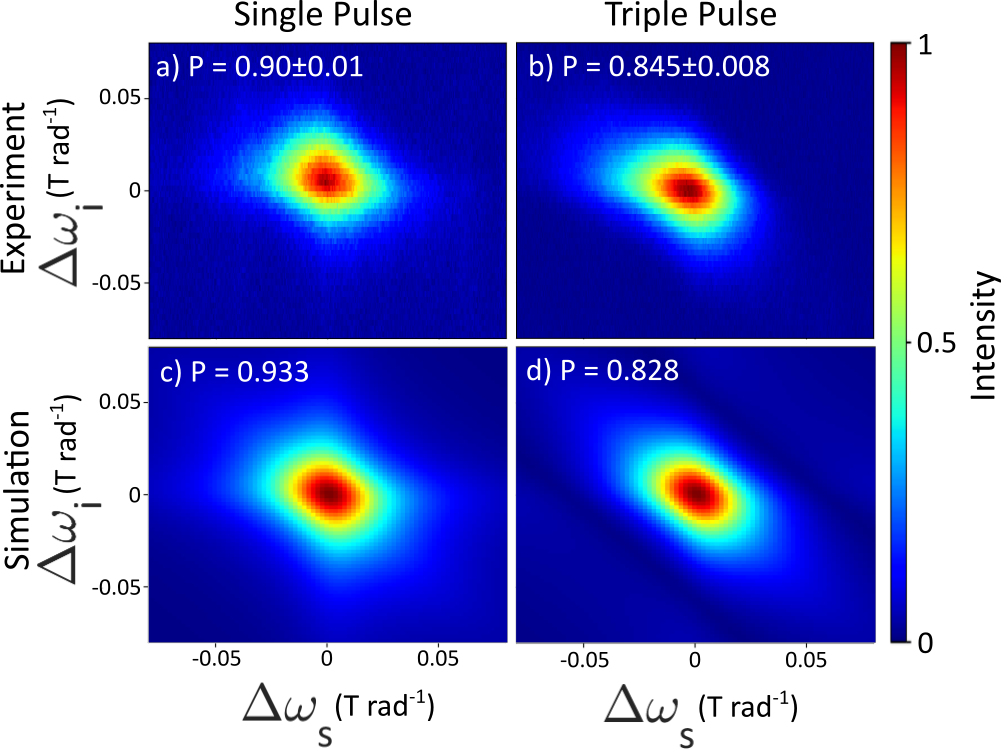}
\caption{Experimentally collected JSIs and calculated Purity P, for a single pulse (a) and triple pulse (b), with $\eta_1 =\eta_2=0.8$ and $\Delta\tau_1 = 20ps$,$\Delta\tau_2 = 40ps$. c) and d) show simulated JSIs for the same single and triple pulse configurations modelled using Eqs.\ref{eq:single} and \ref{eq:tripl}.}
\label{fig:exp}
\end{figure}

Fig.~\ref{fig:exp} presents simulated and experimental data collected for a chosen triple pulse case to show the reduction in purity when a third pulse is added. To establish a baseline with our experimental setup we first recorded JSI for a single pulse as shown in Fig.~\ref{fig:exp}a which agrees with the simulated purity (Fig.~\ref{fig:exp}c). As from the simulation, any triple pulse should show a 0.92 or lower purity when both phases are set to $\pi$ out of phase. Here, $\eta_1$ and $\eta_2$ were set to 0.8 to improve the signal-to-noise ratio of the measured JSI. The result shows an estimated purity of 0.845 which is very close to the simulated purity of 0.828.

\begin{figure}[h]
    \centering
    \includegraphics[width=0.6\linewidth]{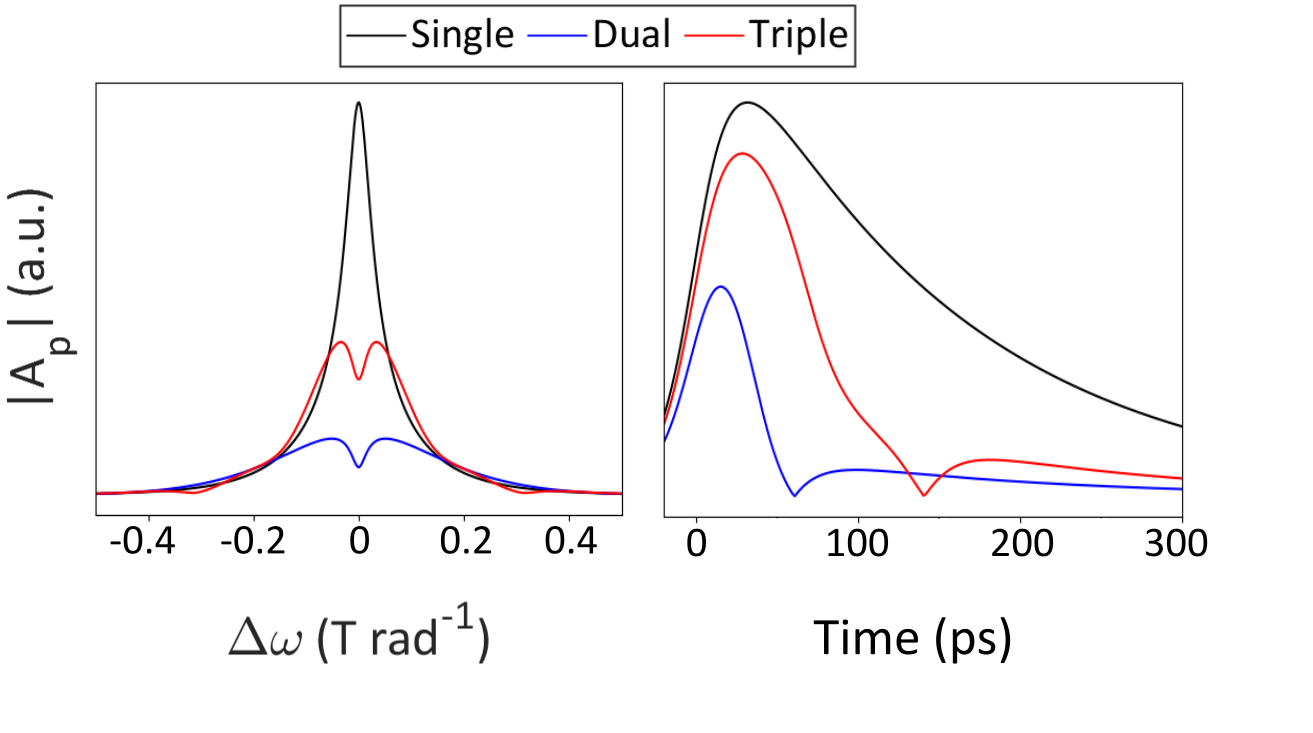}
    \caption{Simulated in-resonator pump fields for single, dual and triple pulse pump, in a) frequency and b) time domain, with dual pulse configuration of $\eta = 0.55$ and $\Delta\tau = 10ps$ and triple pulse configuration of $\eta_1 =\eta_2=0.8$ and $\Delta\tau_1 = 20ps$,$\Delta\tau_2 = 40ps$.}
    \label{fig:Ap}
\end{figure}
Fig.~\ref{fig:Ap} depicts the spectral and temporal pump distribution inside the resonator source ($A_p$), simulated to understand the reduction of purity when more than two pulses are used.  This figure shows that the right configuration of the dual pulse matches best the target spectral shape (Fig.~\ref{fig:sim}) while a triple-pulse adds extra lobes. Adding the second pulse in the dual-pulse configuration significantly reduces temporal pump energy inside the resonator as denoted by the zero-crossing point, thus effectively broadening the pump resonance and improving purity. A third pulse with the same phase as the second adds significantly more energy to the residual energy from the second pulse, hence the effective reduction in purity. While this can be alleviated somewhat by adjusting the phases, the dual pulse still has a quicker energy reduction.
The triple pulse pumps investigated can get close to the desired target pump, but the right dual pulse can fit better, as shown in Fig.\ref{fig:sim}, meaning a dual pulse results in higher purity than any of the tested triple pulses. 
\section{Conclusion}
This work investigated the temporal manipulation of a pulsed pump with greater than two pulses, both through simulation and by using a photonic integrated circuit experiment, on the spectral purity of a ring resonator photon-pairs source. Our results indicate that there was no gain in purity for any pump with more than two pulses for any of the pump profiles investigated. A dual pulse pump provides the best intrinsic match for a specific target pump shape which would result in theoretically maximum purity. An advantage found with the triple pulse is an accurate level of control: with more parameters that are easily adjustable experimentally, the purity can be selected within a large range to any desired value.
However, here the parameters investigated were chosen based on results from the existing literature \cite{Burridge:20, Christensen:18}. In future, a larger parameter space could be more rigorously explored, either corroborating these results or finding outliers with higher purity with comparatively high brightness.

\subsection*{Funding}
B.M.B. acknowledges the support of the EPSRC training grant EP/LO15730/1. The authors acknowledge the support of the EPSRC Quantum Communications Hub (EP/T001011/1) and Quantum Photonic Integrated Circuits (QuPIC) (EP/N015126/1). The authors also include the use of a paid MPW service CORNERSTONE 2 (EP/T019697/1).

\subsection*{Acknowledgments}
L.R would like to thank Hugh Barrett for assisting with the experimental set-up and preliminary investigations.
\subsection*{Disclosures}
The authors declare no conflicts of interest.

\bigskip

\bibliographystyle{unsrt}
\bibliography{sample}

\end{document}